\documentclass[a4paper,11pt]{article}
\usepackage{amsfonts}
\usepackage{amsmath}
\usepackage{amssymb}
\usepackage{graphicx}
\usepackage{array}
\usepackage{booktabs}
\usepackage{multirow}
\usepackage{makecell}
\usepackage{float}
\pdfoutput=1 

\usepackage{mymacros}
\usepackage{chngcntr}
\usepackage{verbatim}
\usepackage{amsthm}
\usepackage{graphics}
\usepackage{xcolor}
\usepackage{mathrsfs}
\usepackage{bbold}
\usepackage{cases}
\usepackage{epsfig}
\usepackage{epstopdf}
\usepackage{hyperref}
\usepackage{amsfonts}
\usepackage{hyperref}
\usepackage{jheppub} 

\usepackage[T1]{fontenc} 

\hypersetup{
	colorlinks=true,
	linkcolor=blue,
	filecolor=blue,
	urlcolor=blue,
	citecolor=blue,
}

\title{\boldmath Spatial regions, chaos bound and its violation}

\author{Zeqiang Wang$^{a}$\footnote{E-mail: zeqwang@hotmail.com}}
\author{Deyou Chen, $^{a}$\footnote{E-mail: deyouchen@hotmail.com}}

\affiliation{$^{a}$School of Science, Xihua University, Chengdu 610039, China}

\abstract{The influence of the angular momentum of the particle on the Lyapunov exponent has been studied. In this paper, we investigate influences of the charge and angular momentum of a particle around non-extremal and extremal Reissner-Nordstr\"om black holes with a scalar hair on the exponent, and find spatial regions where the chaos bound is violated for certain values of the black holes' parameters. The exponent is gotten by calculating the eigenvalue of a Jacobian matrix in a phase space. The violation occurs from the nea-horizon regions to certain distances from the event horizons. When the charge or angular momentum of the particle are fixed at certain values, the spatial regions increase with the increase of the hair parameter's value when the black hole's charge is fixed, and with the increase of the black hole's charge when the hair parameter is fixed. For the extremal black hole, the violation can occur very close to the horizon when the particle's charge is large enough.}

\keywords{Chaos bound, spatial regions, angular momentum.}

\begin{document} 
\maketitle
\flushbottom

\section{Introduction}

Chaos is an important physical phenomenon in nonlinear dynamic systems, and is very sensitive to initial conditions. Its sensitivity is characterized by a Lyapunov exponent. In classical systems, the exponent can be obtained by calculating the eigenvalue of a Jacobian matrix in a phase space and an effective potential approach, respectively. Inspired by AdS/CFT correspondence, out-of-time-order correlators were used to diagnose the chaos in quantum systems \cite{SS1,SS2,SS3,SS4}. In the study of the chaos in the vicinity of the event horizons of black holes, a large number of calculations show that the exponent is related to a system's temperature, $\lambda_L = \frac{2\pi}{\beta}$, where $\lambda_L$ is the exponent and $\beta$ is the inverse temperature of the system. This value is conjectured as a upper bound on the exponent in the work which was gotten by Maldacena, Shenker and Stanford in thermal quantum systems with a large number of degrees of freedom \cite{MSS}. This inspiring work builds a bridge between quantum chaos and black hole physics. After the conjecture was proposed, it immediately attracted much attention and was supported by a lot of work  \cite{PR,MS,CFCZZ,KS,HG,GR1, JKY,RRP1,RRP2,RRP3,RRP4,DMM1,DMM2,CKR1,CKR2,ACO,MPD,HY,GR,CCKM1,BL,MRS,QCL,DG,HMY1,HST1,GLMP,SK1,SK2,SK3,SK4, SK5,SK6,SK7,SK8,SK9,SK10,SK11,LLW,HZ,HZ1,HZ2,HZ3}. For example, it was proved that the bound is saturated by the Sachdev-Ye-Kitaev model, which shows the existence of Einstein gravity dual \cite{AK1,AK2}. On the other hand, Hashimoto and Tanahashi considered a particle moving in the radial direction of a black hole by outward external forces and studied its chaos near the horizon \cite{HT}. They found that the exponent is independent of the strength and species of the external forces and its upper bound is determined by the surface gravity of the horizon, 

\begin{eqnarray}
\lambda_L \leq \kappa.
\label{eq1.1}
\end{eqnarray}

\noindent Considering the relation between the surface gravity and temperature of the black hole, this result is fully consistent with that gotten in \cite{MSS}. 

Recently, the cases of the violation for the bound have been found \cite{ZLL,KG1,KG2,LG1,LG2,GCYW,LTW}. When a particle moves around a black hole, the particle can be in equilibrium due to an external force such as an electromagnetic force. The charge-to-mass ratio and angular momentum of the particle can be adjusted to let it move close to the horizon. Based on this, Zhao et al derived the exponent by calculating the effective potential \cite{ZLL}. To evaluate the violation for the bound in the near-horizon regions of the spherically symmetric black holes, they expanded the exponent at the horizons. They found that the bound is satisfied by the Reissner-Nordstr\"om (RN) and Reissner-Nordstrr\"om anti-de Sitter (RNAdS) black holes, and is violated by a large number of the black holes. In their work, the contribution of the particle's angular momentum was neglected. When this contribution was taken into account, Kan and Gwak studied the exponent in the near-horizon regions and at the certain distances from the horizons of the Kerr-Newman and Kerr-Newman AdS black holes and found that the bound is violated by these black holes \cite{KG1,KG2}. The exponent can be obtained by calculating the eigenvalue of a Jacobian matrix in a phase space. Using this method, Lei and Ge found that the bound is violated by the RN and RNAdS black holes when the contribution of the particle's angular momentum is taken into account \cite{LG2}. In their work, to discuss the exponent in the near-horizon regions, the exponent was also expanded at the horizons. To evaluate the value of the exponent at a certain position from the event horizon, one of us adopted the matrix method and fixed the particle's charge to discuss the influence of the angular momentum of the charged particle on the exponent \cite{GCYW}. In our calculation, the charge of the particle was fixed as a specific value, and the exponent did not expand at the horizon. The violation for the bound was found outside the charged Kiselev black hole. However, the range of the angular momentum and spatial region for the violation were not found.

In this paper, we investigated the influences of the charge and angular momentum of a particle around non-extremal and extremal Reissner-Nordstrr\"om black holes with a scalar hair on the Lyapunov exponent, and find ranges of the angular momentum and charge and spatial regions where the chaos bound is violated. Hairy black holes are important solutions of Einstein’s theory of gravity and modified gravity theories. They have been widely studied mainly in connection with no-hair theorem. The hair is used to describe properties of collapsed stars besides masses, electromagnetic charges and angular momenta. People hope to find hairy black hole solutions with primary scalar hairs. However, most of the solutions are characterized by secondary hairs, and impossible to continuously connect a hair black hole's solution to a black hole's solution with the same mass but no scalar field. On the other hand, hairs might develop in the vicinity of charged black holes. The research of hair black holes is conducive to our deep understanding of the AdS/CFT correspondence \cite{SSG,KAVM,GPSV}. The exponent is derived by calculating the eigenvalue of a Jacobian matrix in a phase space $(r,\pi_r)$.  We first fix the angular momentum at a specific value to discuss the influence of the charge on the exponent and find the range of the charge and spatial region where the bound is violated. Then we fix the charge to discuss the influence of the angular momentum on the exponent and find the spatial region for the violation. The influence of the hair parameter on the exponent is discussed.

The rest is organized as follows. In the next section, we obtain the exponent by calculating the eigenvalue of the Jacobian matrix in the phase space outside the event horizon of the Reissner-Nordstr\"om black hole with a scalar hair. In Section 3, the influences of the charge and angular momentum of the particle around non-extremal and extremal Reissner-Nordstrr\"om-scalar hair black holes on the Lyapunov exponent are investigated, respectively. The spatial regions where the bound is violated are found. The last section is devoted to our conclusions and discussions.

\section{Lyapunov exponent outside RNSH black holes}

The action of general relativity coupled to a Maxwell field and conformally coupled to a scalar field is

\begin{eqnarray}
S=\frac{1}{16\pi G}\int d^4x\sqrt{-g}\left[R- F_{\mu\nu}F^{\mu\nu}- 8\pi G\left( \nabla_\mu \Psi \nabla^\mu \Psi + \frac{R}{6}\Psi^2\right) \right],
\label{eq2.1}
\end{eqnarray}

\noindent where $F_{\mu\nu}$ is the Maxwell field, and $\Psi$ is the scalar field. From the action, Astorino got a static spherically symmetric solution with a scalar hair \cite{MA}. The solution is given by 

\begin{eqnarray}
ds^2 = -f(r) dt^2 + f(r)^{-1}dr^2 + r^2 (d\theta^2  +\sin^2\theta d\phi^2),
\label{eq2.2}
\end{eqnarray}

\noindent with an electromagnetic potential $A_{t}=\frac{Q}{r}$, where

\begin{eqnarray}
f(r) =  1- \frac{2M}{r} + \frac{Q^2+s}{r^2}, \quad \quad
\Psi = \pm \sqrt{\frac{3}{4\pi G}}\sqrt{\frac{s}{Q^2+s}},
\label{eq2.3}
\end{eqnarray}

\noindent $s$ is the scalar hair, $M$ and $Q$ are the mass and charge, respectively. This black hole is referred to as a Reissner-Nordström-scalar hair (RNSH) black hole. The scalar field $\Psi$ exists even in the absence of the electromagnetic field, therefore, the scalar hair is a primary hair. There are two roots for $f(r)=0$,

\begin{eqnarray}
r_{\pm} = M \pm \sqrt{M^2-Q^2-s}, 
\label{eq2.4}
\end{eqnarray}

\noindent  which describe the event horizon ($r_+$) and the inner horizon ($r_-$). The surface gravity is

\begin{eqnarray}
\kappa = \frac{r_+ - r_-}{2r_+^2}.
\label{eq2.5}
\end{eqnarray}

\noindent When the inner and event horizons coincide with each other, there is $M^2=Q^2+s$ and the surface gravity disappears.

A Lyapunov exponent denotes the average rate of contraction and expansion of two adjacent orbits in a classical phase space. It can gotten by the effective potential method and matrix method. Here we first review the matrix method to derive the exponent \cite{LG1,LG2,CMBWZ,PP1,PP2,PP3}. From the equations of motion of a particle, 

\begin{eqnarray}
\frac{dX_i(t)}{dt} = F_i(X^j),
\label{eq2.1.2}
\end{eqnarray}

\noindent we use a certain orbit and linearize the equations,

\begin{eqnarray}
\frac{d\delta X_i(t)}{dt} =K_{ij}(t)\delta X_j(t),
\label{eq2.1.3}
\end{eqnarray}

\noindent where $K_{ij}(t) $ is a linear stability matrix defined by $K_{ij}(t) = \left.\frac{\partial F_i}{\partial X_j}\right|_{X_i(t)}$. The solution of Eq. (\ref{eq2.1.3}) satisfies 

\begin{eqnarray}
\delta X_i(t) = L_{ij}(t)\delta X_j(0),
\label{eq2.145}
\end{eqnarray}

\noindent where $L_{ij}(t)$ is a evolution matrix and obeys $ \dot{L}_{ij}(t)=K_{im} L_{mj}(t)$ and $L_{ij}(0) = \delta_{ij}$. The exponent is obtained from the determination of the eigenvalues of the matrix $L_{ij}$. It can be written as 

\begin{eqnarray}
\lambda = \lim_{t \to \infty} \frac{1}{t}ln\left(\frac{L_{jj}(t)}{L_{jj}(0)}\right).
\label{eq2.1.5}
\end{eqnarray}

\noindent The positive exponent implies a chaos in the system. When we focus our attention on a circular orbit in an equatorial plane of a black hole, the classical phase space is expressed as $(p_r, r)$. Let's further linearize the equation of motion with $X_i(t) = (p_r, r)$ about the orbit of constant $r$, and then get $K_{11}=K_{22}=0$,  $K_{12}=\frac{d}{dr}\left(\dot{t}^{-1}\frac{\delta \mathcal{L}}{\delta r}\right)$ and $K_{21}=-(\dot{t}g_{rr})^{-1}$, where $\mathcal{L}$ is the Lagrangian for geodesic motion. Thus, the exponent is gotten as 

\begin{eqnarray}
\lambda = \sqrt{K_{12}K_{21}}.
\label{eq2.1.6}
\end{eqnarray}

To derive the exponent via the matrix method, we first find the motion equation of a particle with charge $q$ around the equatorial plane of the black hole. The Lagrangian of the particle is

\begin{eqnarray}
\mathcal{L} = \frac{1}{2}\left(-f\dot{t}^2+\frac{\dot{r}^2}{f} +r^2\dot{\phi}^2\right) -qA_t\dot{t},
\label{eq2.6}
\end{eqnarray}

\noindent where $f=f(r)$, $\dot{x^{\mu}} = \frac{dx^{\mu}}{d\tau}$ and $\tau$ is a proper time. We use the definition of  generalized momenta $ \pi_{\mu}=\frac{\partial\mathcal{L}}{\partial\dot{x}}$ and get

\begin{eqnarray}
\pi_t = -f\dot{t} -qA_t=-E, \quad\quad \pi_r = \frac{\dot{r}}{f}, \quad\quad \pi_{\phi} = r^2 \dot{\phi}=L.
\label{eq2.7}
\end{eqnarray}

\noindent Due to Killing vectors in the black hole background, $E$ and $L$ are constants in the above equation and denote an energy and angular momentum, respectively. The particle's Hamiltonian is 

\begin{eqnarray}
H = \frac{-(\pi_{t}+qA_{t})^2+\pi_r^2f^2+ \pi^2_{\phi}r^{-2}f}{2f}.
\label{eq2.8}
\end{eqnarray}

\noindent Equations of motion of the particle are obtained from the Hamiltonian, which are

\begin{eqnarray}
\dot{t} &=& \frac{\partial H}{\partial \pi_t}=-\frac{\pi_t+qA_t}{f}, \quad  \dot{\pi_t}= -\frac{\partial H}{\partial t} =0 ,
\quad \dot{r} = \frac{\partial H}{\partial \pi_r}= \pi_rf, \nonumber\\
\dot{\pi_r} &=& -\frac{\partial H}{\partial r} =-\frac{1}{2}\left[\pi^2_r f^{\prime} -\frac{2qA^{\prime}_t(\pi_t+qA_t)}{f}+\frac{(\pi_t+qA_t)^2f^{\prime}}{f^2} -\pi^2_{\phi}(r^{-2})^{\prime}\right], \nonumber\\
\dot{\phi} &=& \frac{\partial H}{\partial \pi_{\phi}}= \frac{\pi_{\phi}}{r^2}, \quad  \dot{\pi_\phi}= -\frac{\partial H}{\partial \phi} =0,
\label{eq2.9}
\end{eqnarray}

\noindent where "$\prime$" denotes a derivative with respect to $r$. We derive the exponent in a phase space $(r, \pi_r)$. Therefore, a radial equation at a coordinate time $t$ should be first obtained. From the equations of motion, we get

\begin{eqnarray}
\frac{dr}{dt} &=& \frac{\dot{r}}{\dot{t}} =-\frac{\pi_rf^2}{\pi_t+qA_t}, \nonumber\\
\frac{d\pi_r}{dt} &=& \frac{\dot{\pi_r}}{\dot{t}} = -qA^{\prime}_t +\frac{1}{2}\left[\frac{\pi^2_r ff^{\prime}}{\pi_t+qA_t}+\frac{(\pi_t+qA_t)f^{\prime}}{f} -\frac{\pi^2_{\phi}(r^{-2})^{\prime}f}{\pi_t+qA_t}\right].
\label{eq2.10}
\end{eqnarray}

\noindent There is a normalization condition for the four-velocity of a particle, $ g_{\mu\nu}\dot{x}^{\mu}\dot{x}^{\nu}=\eta$, where $\eta =0$ describes a case of a massless particle and $\eta =-1$ denotes a case of a massive particle. When a charged particle is considered, the normalization yields $\pi_t+qA_t=-\sqrt{f(1+ \pi_r^2f + \pi_{\phi}^2r^{-2})}$. We use this constrain and define $F_1= \frac{dr}{dt}$, $F_2=\frac{d\pi_r}{dt}$. Then Eq. (\ref{eq2.10}) is rewritten as  

\begin{eqnarray}
F_1 &=& \frac{\pi_rf^2}{\sqrt{F(1+ \pi_r^2f + \pi_{\phi}^2r^{-2})}}, \nonumber\\
F_2 &=& -qA^{\prime}_t -\frac{(2\pi^2_r f+1)f^{\prime}}{2\sqrt{f(1+ \pi_r^2f + \pi_{\phi}^2r^{-2})}} -\frac{\pi^2_{\phi}(r^{-2}f)^{\prime}}{2\sqrt{f(1+ \pi_r^2f + \pi_{\phi}^2r^{-2})}}.
\label{eq2.11}
\end{eqnarray}

In the phase space $(r, \pi_r)$, the elements of the matrix is defined by 

\begin{eqnarray}
K_{11} = \frac{\partial F_1}{\partial r} , \quad\quad
K_{12} = \frac{\partial F_1}{\partial \pi_r} ,\quad\quad
K_{21} = \frac{\partial F_2}{\partial r} ,\quad\quad
K_{22} = \frac{\partial F_2}{\partial \pi_r}.
\label{eq2.12}
\end{eqnarray}

\noindent When the particle is in equilibrium in a circular orbit, its trajectory satisfies a constraint $\pi_r=\frac{d\pi_r}{dt} =0$. Then locations of equilibrium orbits are derived by using the constraint and Eq. (\ref{eq2.11}). The exponent of the chaos of the charged particle in an equilibrium orbit is obtained by calculating the eigenvalue of the matrix, which is identifies

\begin{eqnarray}
\lambda^2 = \frac{1}{4}\left[\frac{f^{\prime}+\pi_{\phi}^2(r^{-2}f)^{\prime}}{1+\pi_{\phi}^2r^{-2}} \right]^2 -\frac{1}{2}f\frac{f^{\prime\prime}+\pi_{\phi}^2(r^{-2}f)^{\prime\prime}}{1+\pi_{\phi}^2r^{-2}}
-\frac{qA_t^{\prime\prime}f^2}{\sqrt{f(1+\pi_{\phi}^2r^{-2})}}.
\label{eq2.13}
\end{eqnarray}

\noindent At the event horizon, we get $f(r_+)=0$ and 

\begin{eqnarray}
\lambda^2 =\frac{1}{4}(f^{\prime})^2=\kappa^2,
\end{eqnarray}

\noindent which is saturated at the horizon. This result is consistent with that obtained in the spherically symmetric black holes \cite{ZLL,LG2}. In the derivation of Eq. (\ref{eq2.13}), we fixed the particle’s charge in the calculation process, and didn't expand the  exponent at the event horizon. When the charge is expressed in terms of $\pi_{\phi}$, $r$ and $A_t$, the expression of the exponent in \cite{LG2} can be easily recovered. It is clearly that the exponent is affected by the angular momentum and charge of the particle. When the angular momentum is neglected and the expression of the charge is adopted, the exponent obtained in \cite{ZLL} is gotten. 

\section{Bound on Lyapunov exponent outside RNSH black holes}

When a charged particle moves around a black hole, the charge-to-mass ratio and angular momentum of the particle can be adjusted to make it in equilibrium at a certain position from the event horizon. In this section, we fix the particle's charge (or angular momentum) to investigate the influences of the angular momentum (or the charge) on the Lyapunov exponent, and to find spatial regions where the chaos bound is violated. The values of the exponent and surface gravity are gotten by numerically calculating Eqs. (\ref{eq2.5}) and (\ref{eq2.13}), respectively. Without loss of generality, we order $M=1$ in this paper.

\begin{figure}[h]
	\centering
	\includegraphics[width=10cm,height=6cm]{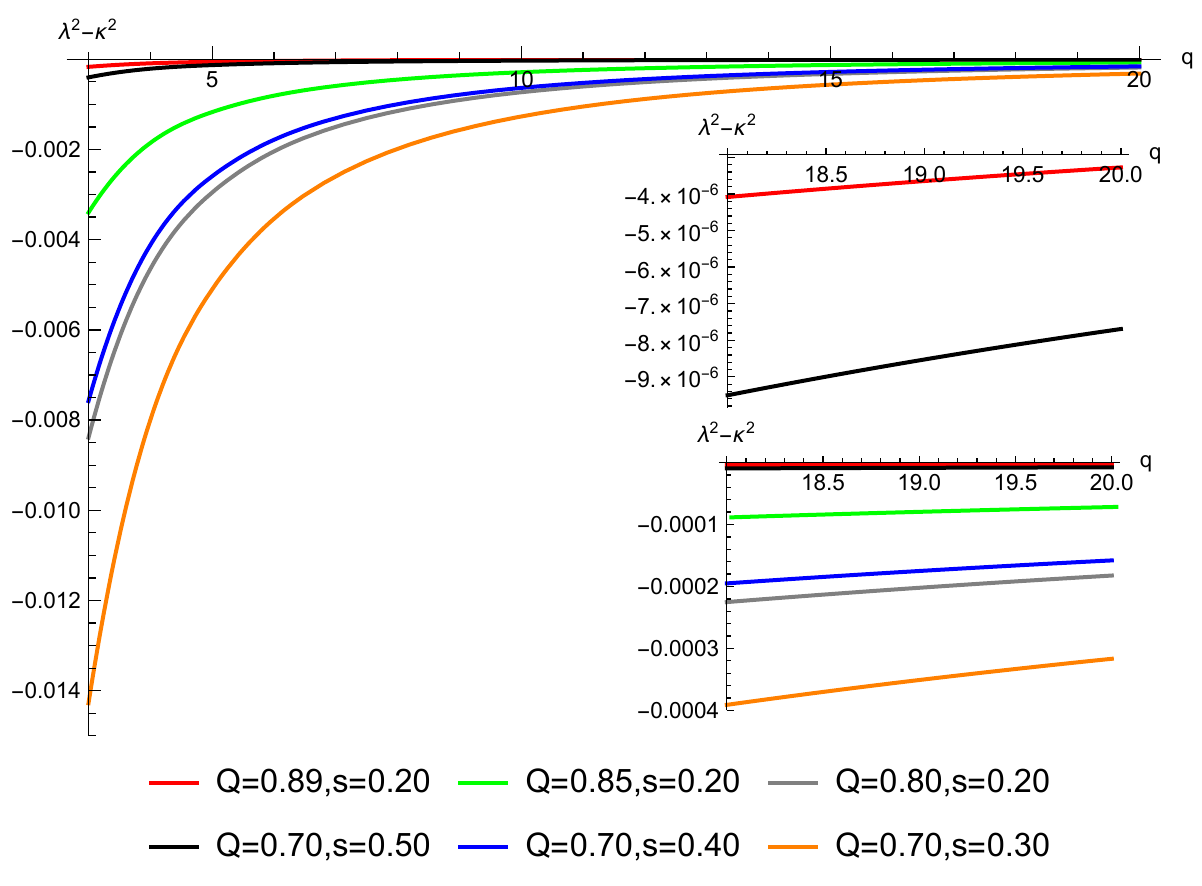}
	\caption{The influence of the charge of the particle around the non-extremal RNSH black hole on the Lyapunov exponent, where $L = 0$. }
\end{figure}

\begin{figure}[h]
	\centering
	\includegraphics[width=10cm,height=6cm]{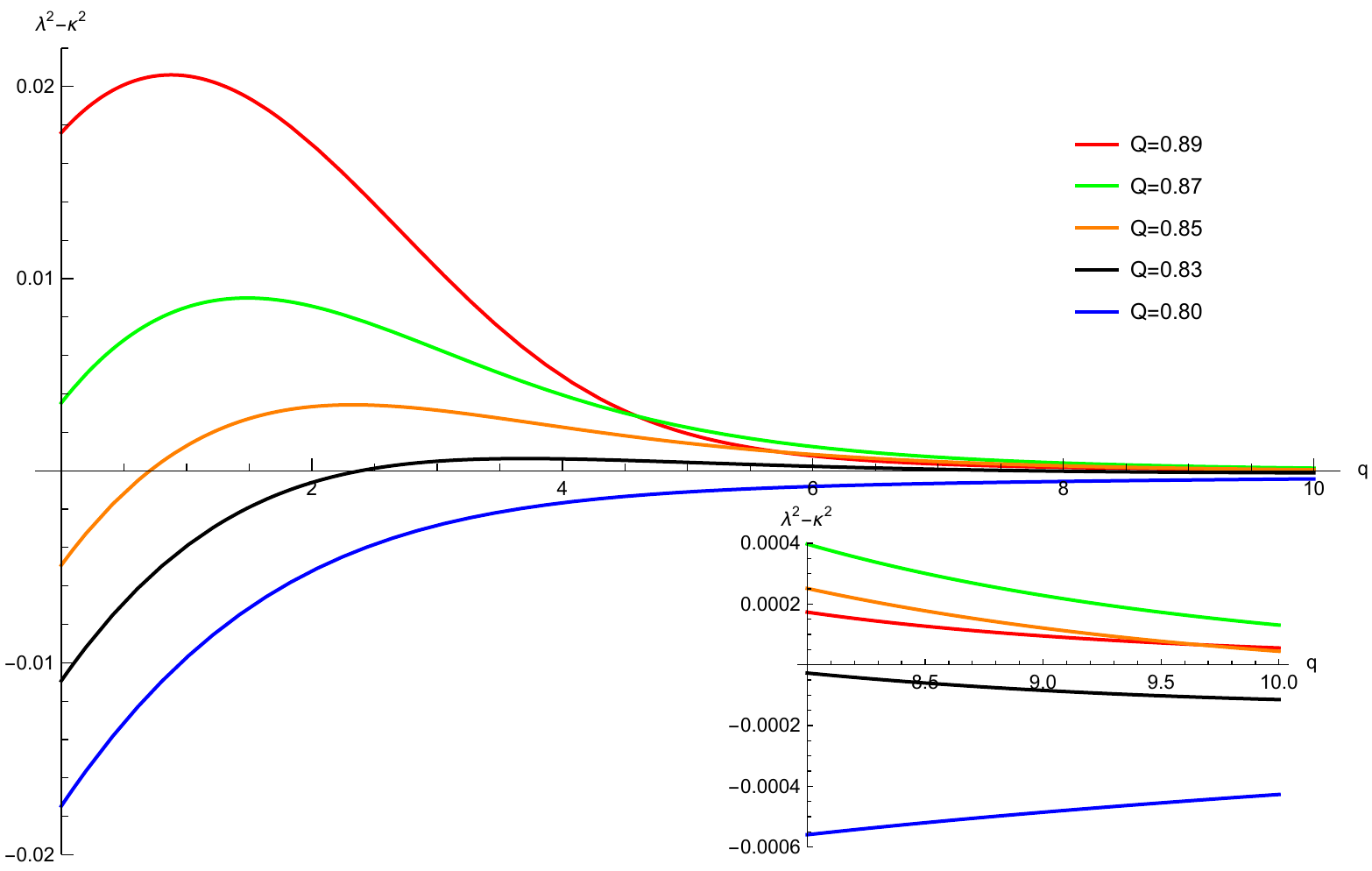}
	\caption{The influence of the charge of the particle around the non-extremal RNSH black hole on the Lyapunov exponent, where $L = 5$ and $s=0.20$. The chaos bound is violated in the range  $0.01<q<20.67$ (the corresponding spatial region is $1.00327r_+ <r_0 < 2.28201r_+$, where $r_+ = 1.08888$) when $Q=0.89$, in the range $0.1<q<14.35$ ($ 1.01477r_+<r_0<1.85038r_+$, where $r_+ = 1.20761$) when $Q=0.87$, in the range $0.71<q<11.97$ ($1.03233r_+<r_0<1.61754r_+$, where  $r_+ = 1.27839$) when $Q=0.85$, and in the range $2.39<q<7.66$ ($1.07181r_+<r_0<1.32367r_+$, where $r_+ = 1.33332$) when $Q=0.83$. There is no violation  when $Q=0.80$.}
\end{figure}

\begin{figure}[h]
	\centering
	\includegraphics[width=10cm,height=6cm]{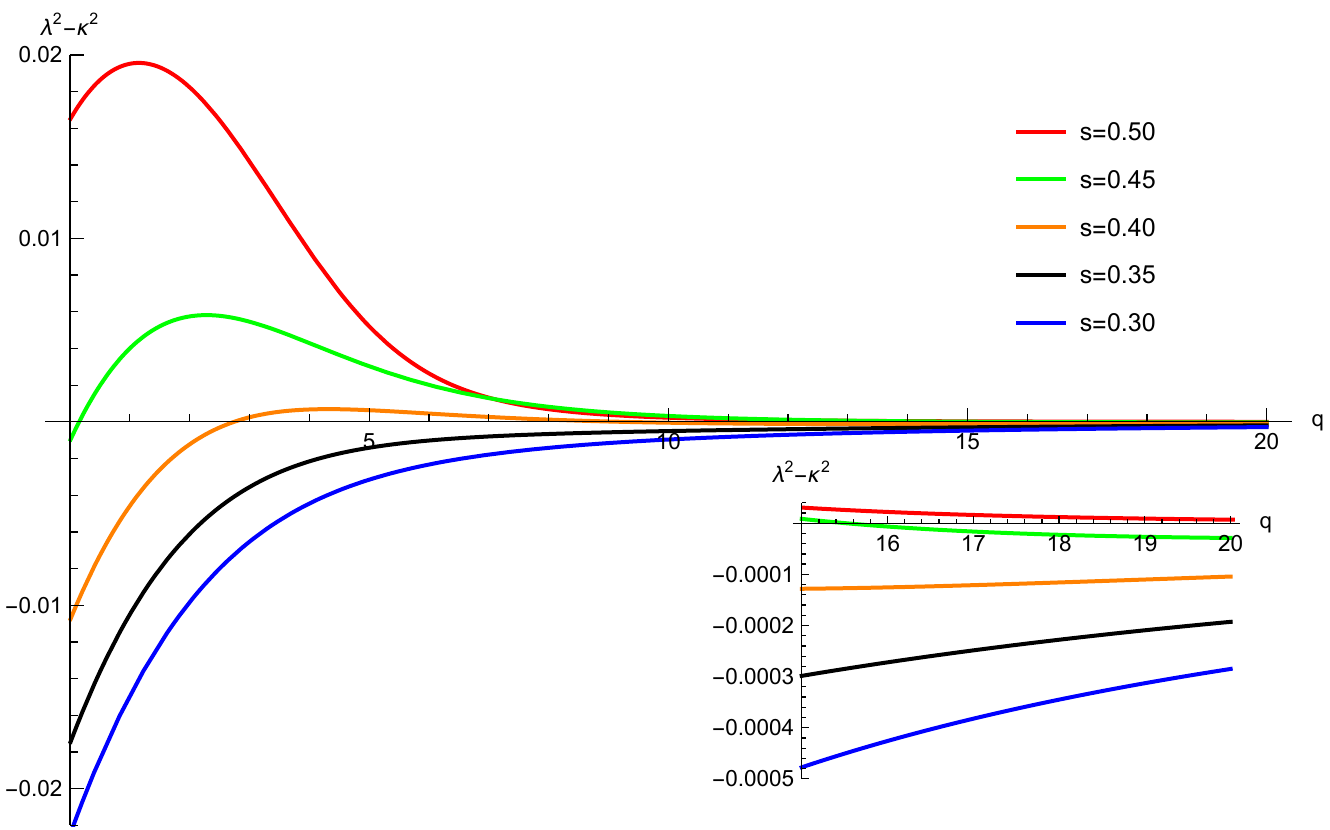}
	\caption{The influence of the charge of the particle around the non-extremal RNSH black hole on the Lyapunov exponent, where $L = 5$ and $Q=0.70$. The chaos bound is violated in the range $0.01<q<24.86$ ($ 1.00041r_+<r_0<1.97703r_+$, where $ r_+ = 1.10000$) when $s=0.50$, in the range $0.14<q<15.51$ ($1.02205r_+<r_0<1.81672r_+$, where $r_+ = 1.24495$) when $s=0.45$, and in the range $2.74<q<9.25$ ($1.06960r_+<r_0<1.33271r_+$, where $r_+ = 1.33166$) when $s=0.40$. There is no violation when $s=0.35$ and $s=0.30$.}
\end{figure}

The influence of the charge of the particle around the non-extermal RNSH black hole on the exponent is plotted in Figure 1 - Figure 3 when the angular momentum is fixed at $L=0$ and $L=5$, respectively. In Figure 1, we calculate the exponent when the black hole's parameters are certain values and the angular momentum is zero, and find there is no violation for these values. When the particle's charge is small, equilibrium orbits don't exist. The influence of the particle's charge on the exponent is described in Figure 2 and 3 when the angular momentum is fixed at $L=5$. In Figure 2, the range of the charge $q$ and spatial region where the bound is violated increase with the increase of the black hole's charge. There is no violation when $Q =0.80$. When the particle's charge is less than $1$, equilibrium orbits exist, and the violation occurs for $Q=0.89$, $Q=0.87$ and $Q=0.85$. In Figure 3, the range of the particle's charge and spatial region for the violation increase with the increase of the hair parameter's value when the black hole's charge is fixed at $Q=0.70$. A large value of the hair parameter is more likely to cause a violation. In the above figures, the angular momentum plays an important role in the violation of the bound. When the angular momentum is very small (for example, $ L=0 $), there is no violation regardless of how the particle's charge, black hole's charge and hair parameter change. Only when the angular momentum is relatively large, the bound can be violated for this black hole. Therefore, we will discuss the case where the angular momentum is not zero in the following.

\begin{figure}[h]
	\centering
	\includegraphics[width=10cm,height=6cm]{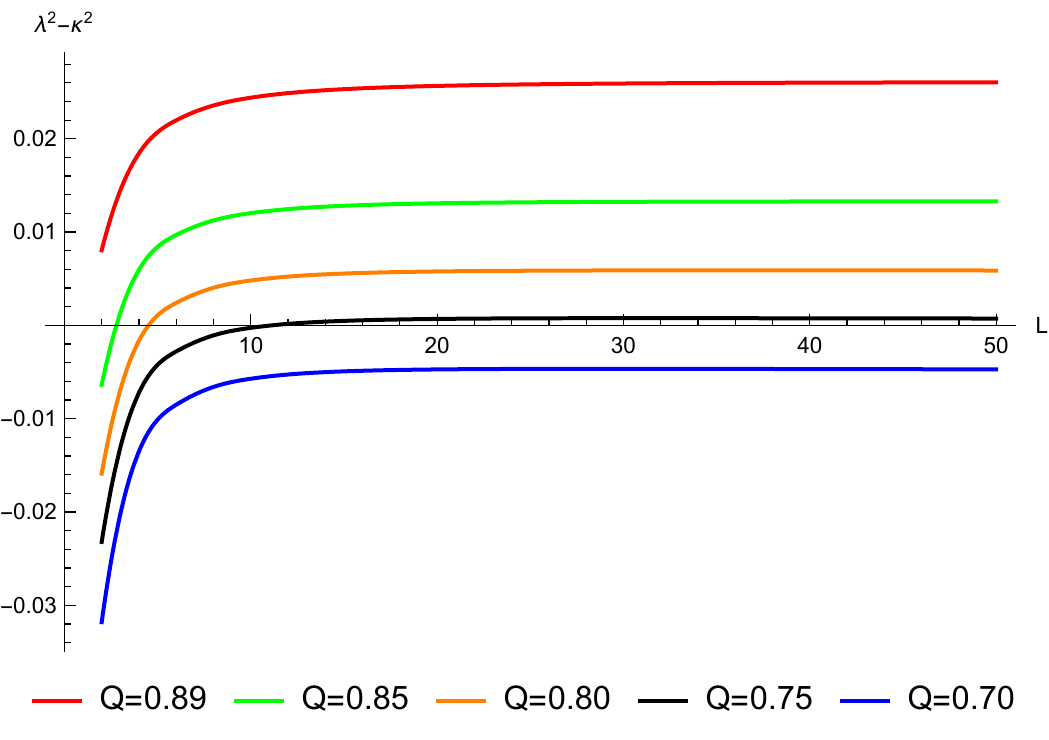}
	\caption{The influence of the angular momentum of the particle around the non-extremal RNSH black hole on the Lyapunov exponent, where $q=0.90$ and $s=0.20$. The chaos bound is violated in the range $L>1.38$ (the corresponding spatial region is $1.85103r_+<r_0<1.98131r_+$, where $r_+ = 1.08888$) when $Q=0.89$, in the range $L>2.70$ ($1.61888r_+<r_0<1.72215r_+$, where $r_+ = 1.20761$) when $Q=0.87$, in the range $L>4.58$ ($1.57125r_+<r_0<1.67105r_+$, where $r_+ = 1.27839$) when $Q=0.85$, and in the range $L>11.09$ ($1.57763r_+<r_0<1.64029r_+$, where $r_+ = 1.33332$) when $Q=0.83$. There is no violation when $Q=0.80$.}
\end{figure}

\begin{figure}[h]
	\centering
	\includegraphics[width=10cm,height=6cm]{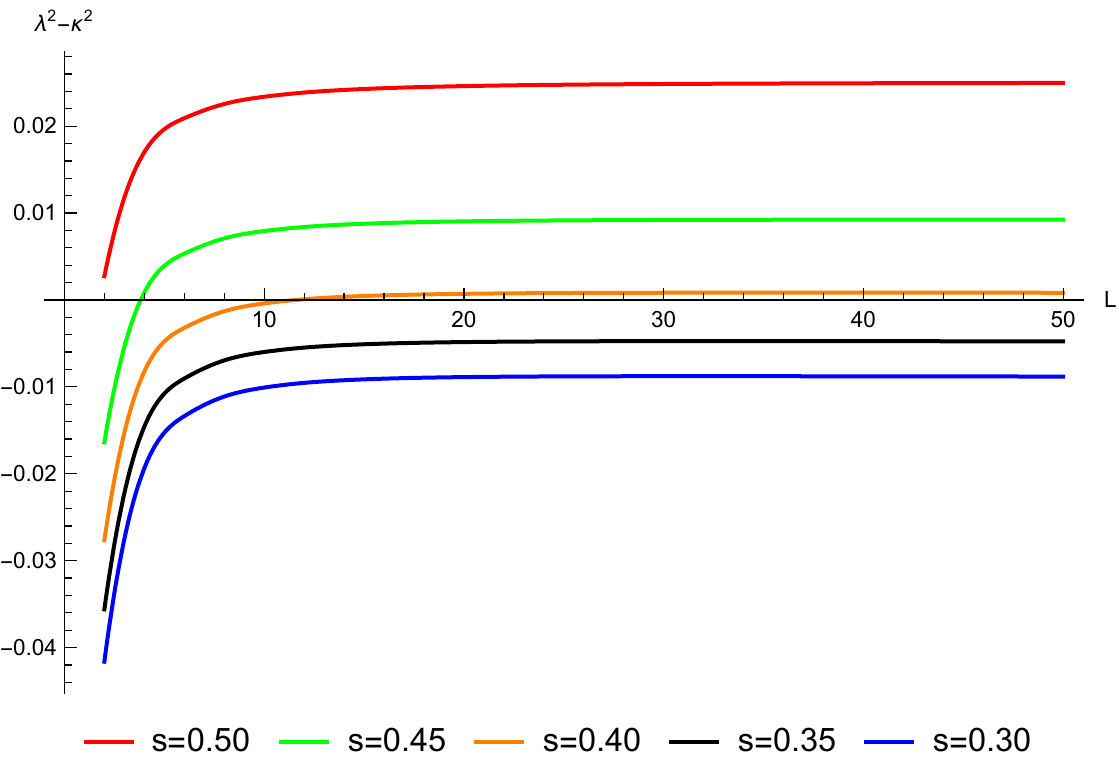}
	\caption{The influence of the angular momentum of the particle around the non-extremal RNSH black hole on the Lyapunov exponent, where $q=0.90$ and $Q=0.70$. The chaos bound is violated in the range $L>1.85$ ($ 1.83590r_+<r_0<2.13661r_+$, where $r_+ = 1.10000$) when $s=0.50$, in the range $L>3.78$ ($1.64460r_+<r_0<1.69337r_+$, where $ r_+ = 1.24495$) when $s=0.45$, and in the range $L>11.56$ ($1.59273r_+<r_0<1.64115r_+$, where $r_+ = 1.33166$) when $s=0.40$. There is no violation when $s=0.35$ and $s=0.30$.}
\end{figure}

\begin{figure}[h]
	\centering
	\includegraphics[width=10cm,height=6cm]{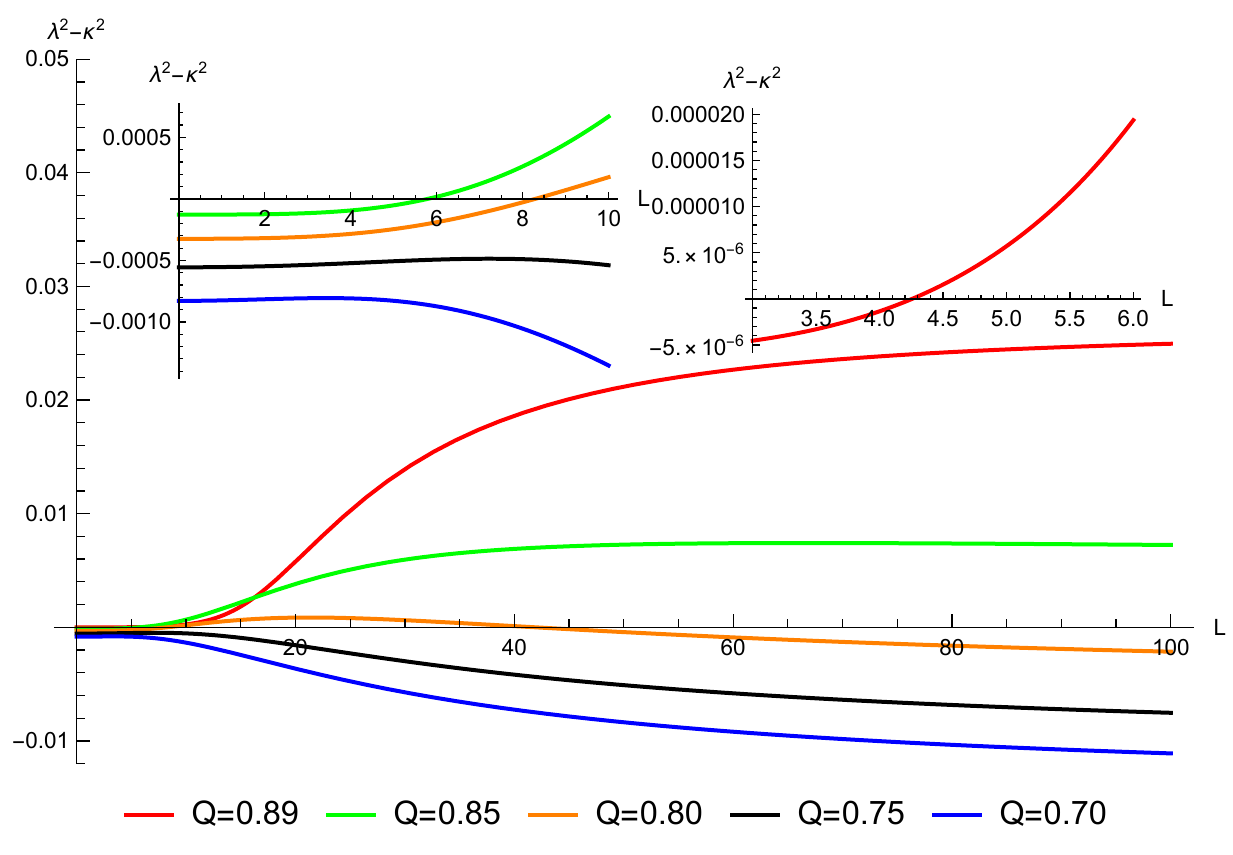}
	\caption{The influence of the angular momentum of the particle around the non-extremal RNSH black hole on the Lyapunov exponent, where $q=15$ and $s=0.20$. The chaos bound is violated in the range $L>4.25$ ($ 1.00463r_+<r_0<1.84862r_+$, where $r_+ = 1.08888$) when $Q=0.89$, in the range $L>5.79$ ($1.02301r_+<r_0<1.67098r_+$, where $r_+ = 1.27839$) when $Q=0.85$, and in the range $8.30<L<41.86$ ($ 1.05939r_+<r_0<1.34503r_+$, where $r_+ = 1.40000$) when $Q=0.80$. There is no violation when $Q=0.75$ and $Q=0.70$.}
\end{figure}

\begin{figure}[h]
	\centering
	\includegraphics[width=10cm,height=6cm]{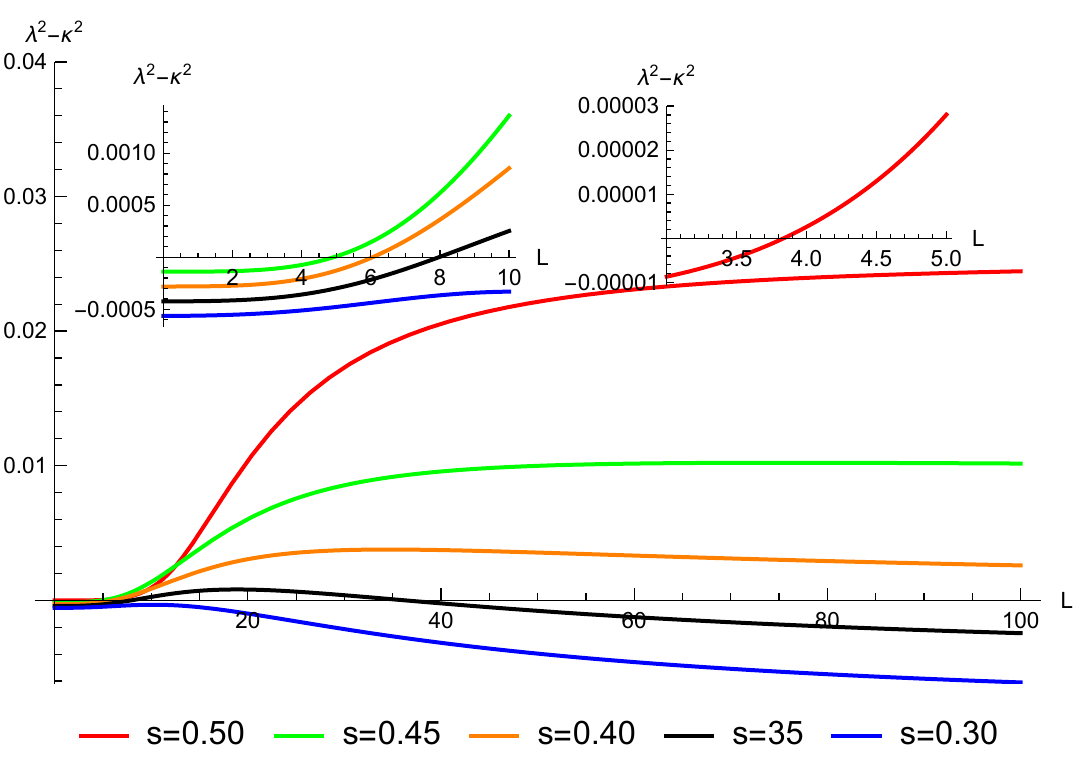}
	\caption{The influence of the angular momentum of the particle around the non-extremal RNSH black hole on the Lyapunov exponent, where $q=15$ and $Q=0.70$. The chaos bound is violated in the range $L>3.84$ ($1.00698r_+<r_0<1.83414r_+$, where $r_+ = 1.10000$) when $s=0.50$, in the range $L>4.92$ ($1.02286r_+<r_0<1.69188r_+$, where $r_+ = 1.24495$) when $s=0.45$, in the range $L>6.05$ ($1.04015r_+<r_0<1.63982r_+$, where $r_+ = 1.33166$) when $s=0.4$, and in the range $7.97<L<36.21$ ($1.06953r_+<r_0<1.34309r_+$, where $r_+ = 1.40000$) when $s=0.35$. There is no violation when $s=0.30$.}
\end{figure}

The influence of the angular momentum of the particle around the non-extermal RNSH black hole on the exponent is plotted in Figure 4 - Figure 7 by fixing $q = 0.9$ and $q = 15$, respectively. Figure 4 shows that the angular momentum's range and spatial region increase with the increase of the black hole's charge when $q=0.9$ and $s=0.2$. There is no violation when the black hole's charge less than a certain value. For example, there is no violation when $Q=0.80$. In Figure 5, we observe that the angular momentum's range and spatial region decrease with the decrease of the hair parameter's value. There is no violation when $s=0.35$ and $s=0.30$. Although the ranges of the angular momentum in these two figures are very large, the corresponding spatial regions are not large. In \cite{ZLL,LG1,LG2}, the authors did not directly give the charge-to-mass ratio of the particle, but gave a general expression about the charge. In \cite{KG1,KG2}, the authors found the violation for the bound when the charge-to-mass ratio is greater than 1. In above discussion, there is also a violation when the ratio is less than 1. Therefore, the violation found here does not require a weak gravity conjecture. We will discuss the case where the ratio is greater than 1 in the following.

In Figure 6, the angular momentum's range and spatial region increase with the increase of the black hole's charge. When the charge is less than a certain value, there is no violation. When $Q=0.89$, $s=0.20$ and the angular momentum approaches 4.25, there is $r_0\to 1.00463r_+$, which indicates that the equilibrium orbit is in the near-horizon region. Therefore, the bound is violated from the near-horizon region to a certain distance from the horizon. The similar case also occurs in Figure 7, where $L \to 3.84$ and $r_0\to 1.00698r_+$ when $Q=0.70$ and $s=0.50$. In Figure 7, the angular momentum's range and spatial region increase with the increase of the hair parameter's value. When the hair parameter is less than a certain value, there is no violation. Comparing Figure 4 and 5 with Figure 6 and 7, we find that for the same black hole's parameters, the spatial regions are relatively large when the particle's charge is large. The reason for this phenomenon is that the electromagnetic force applied to the particle increases with the increase of its charge. The particle with large charge is farther away from the horizon than that with less charge.

\begin{figure}[h]
	\centering
	\includegraphics[width=10cm,height=6cm]{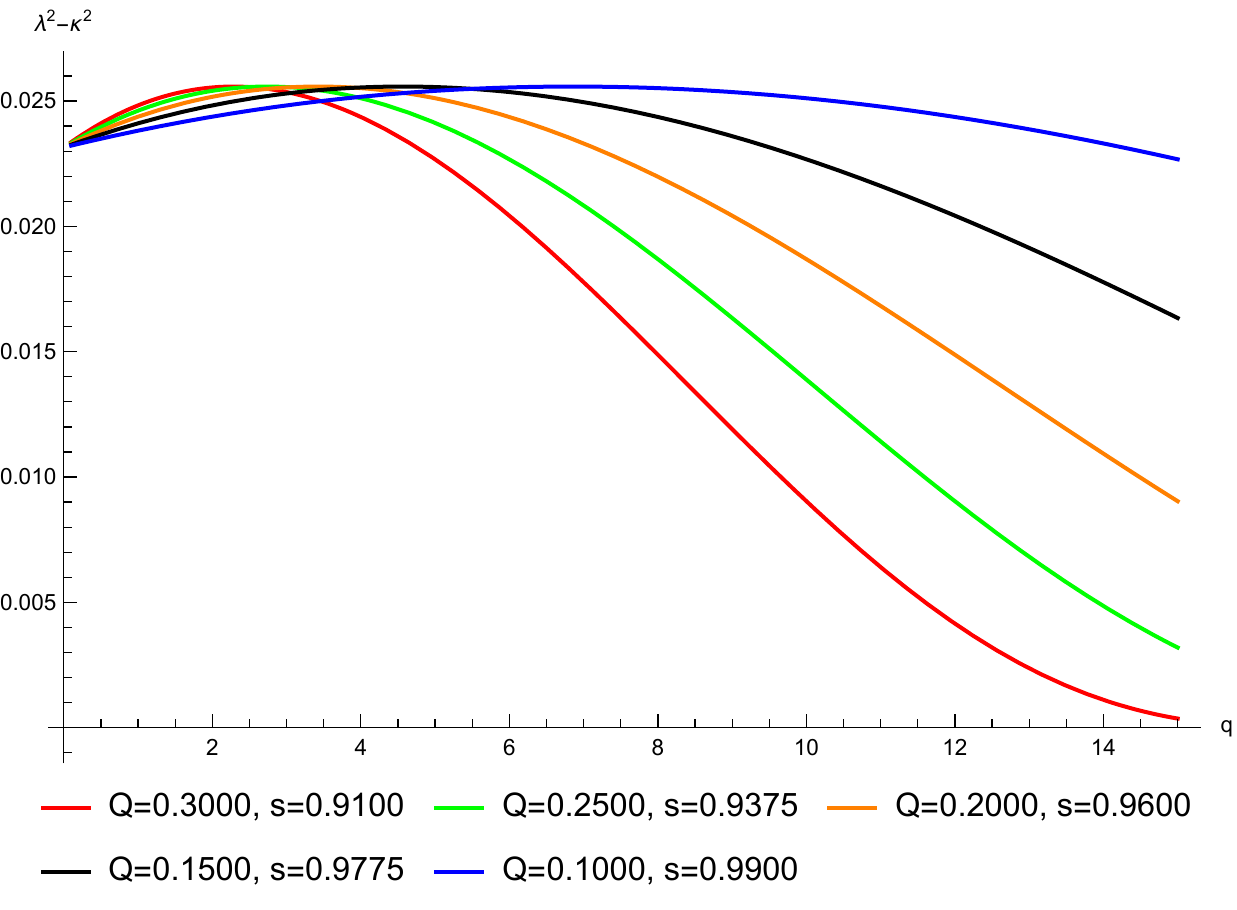}
	\caption{The influence of the charge of the particle around the extremal RNSH black hole on the Lyapunov exponent, where $L=5$. The chaos bound is violated in the range $0<q<16.99$ ($1.000206r_+<r_0<2.192235r_+$) when $Q=0.3000$ and $s=0.9100$, in the range $0<q<20.39$ ($1.000155r_+<r_0<2.192235r_+$) when $Q=0.2500$ and $s=0.9375$, in the range $0<q<25.49$ ($1.000104r_+<r_0<2.192235r_+$) when $Q=0.2000$ and $ s=0.9600$, in the range $0<q<33.99$ ($1.000053r_+<r_0<2.192235r_+$) when $Q=0.1500$ and $ s=0.9775$, and in the range $0<q<50.98$ ($1.000002r_+<r_0<2.192235r_+$) when $Q=0.1000$ and $ s=0.9900$.}
\end{figure}

\begin{figure}[h]
	\centering
	\includegraphics[width=10cm,height=6cm]{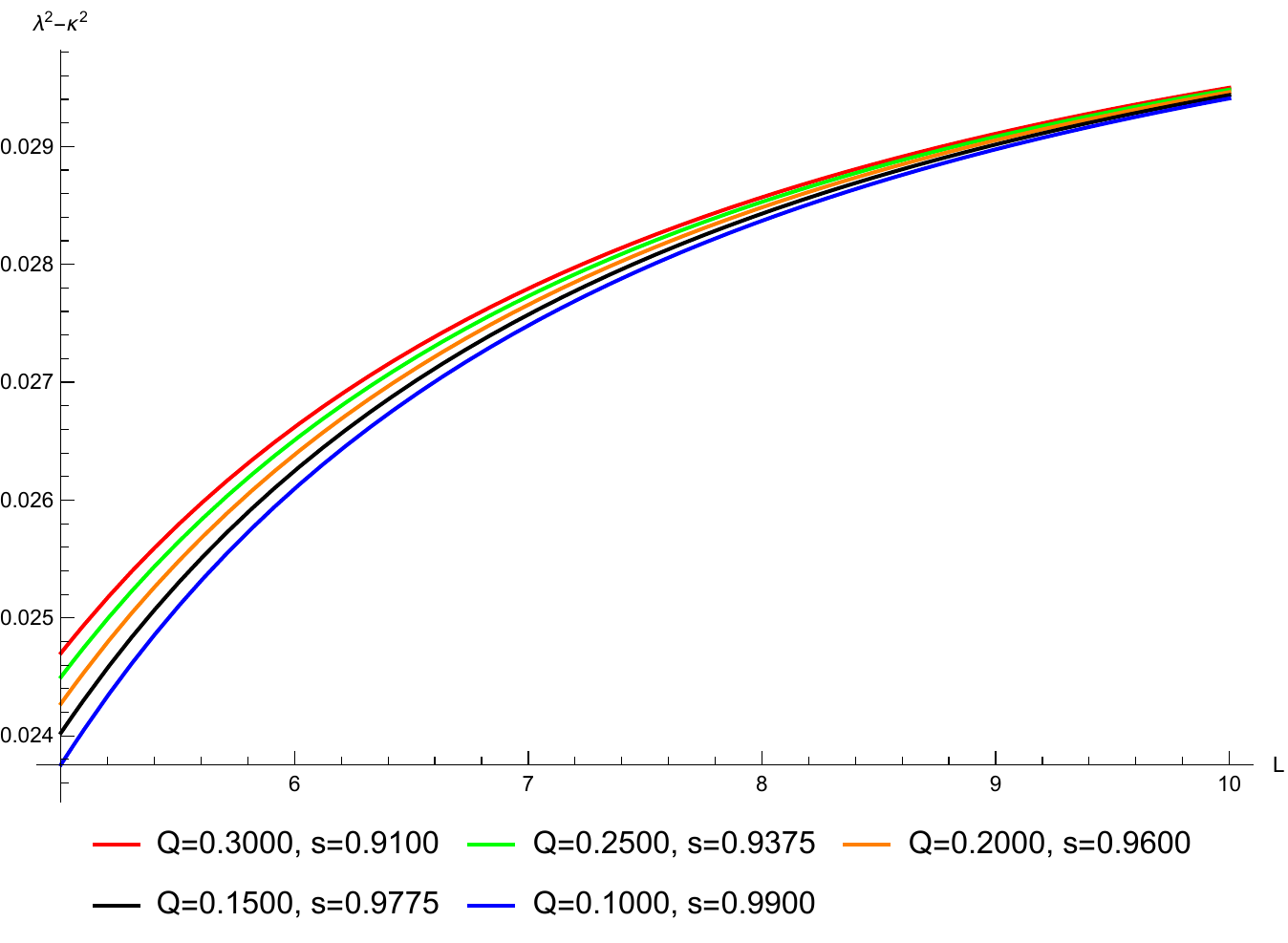}
	\caption{The influence of the angular momentum of the particle around the extremal RNSH black hole on the Lyapunov exponent, where $q=0.9$. The chaos bound is violated in the range $L > 2.34$ $(1.98144r_+ < r_0 < 3.73791r_+)$ when $Q = 0.3000$ and $s = 0.9100$, in the range $ L > 2.43$ $(1.98718r_+ < r_0 < 3.78975r_+)$
when $Q = 0.2500$ and $s = 0.9375$, in the range $L > 2.51$ $(1.99183r_+ < r_0 < 3.83103r_+)$ when $Q = 0.2000$ and $s = 0.9600$, in the range $ L > 2.59$ $(1.99542r_+ < r_0 < 3.86573r_+)$ when $Q = 0.1500$ and $s = 0.9775$, $L > 2.67$ $(1.99797r_+ < r_0 < 3.89825r_+)$ when $Q = 0.1000$ and $s = 0.9900$. }
\end{figure}

\begin{figure}[h]
	\centering
	\includegraphics[width=10cm,height=6cm]{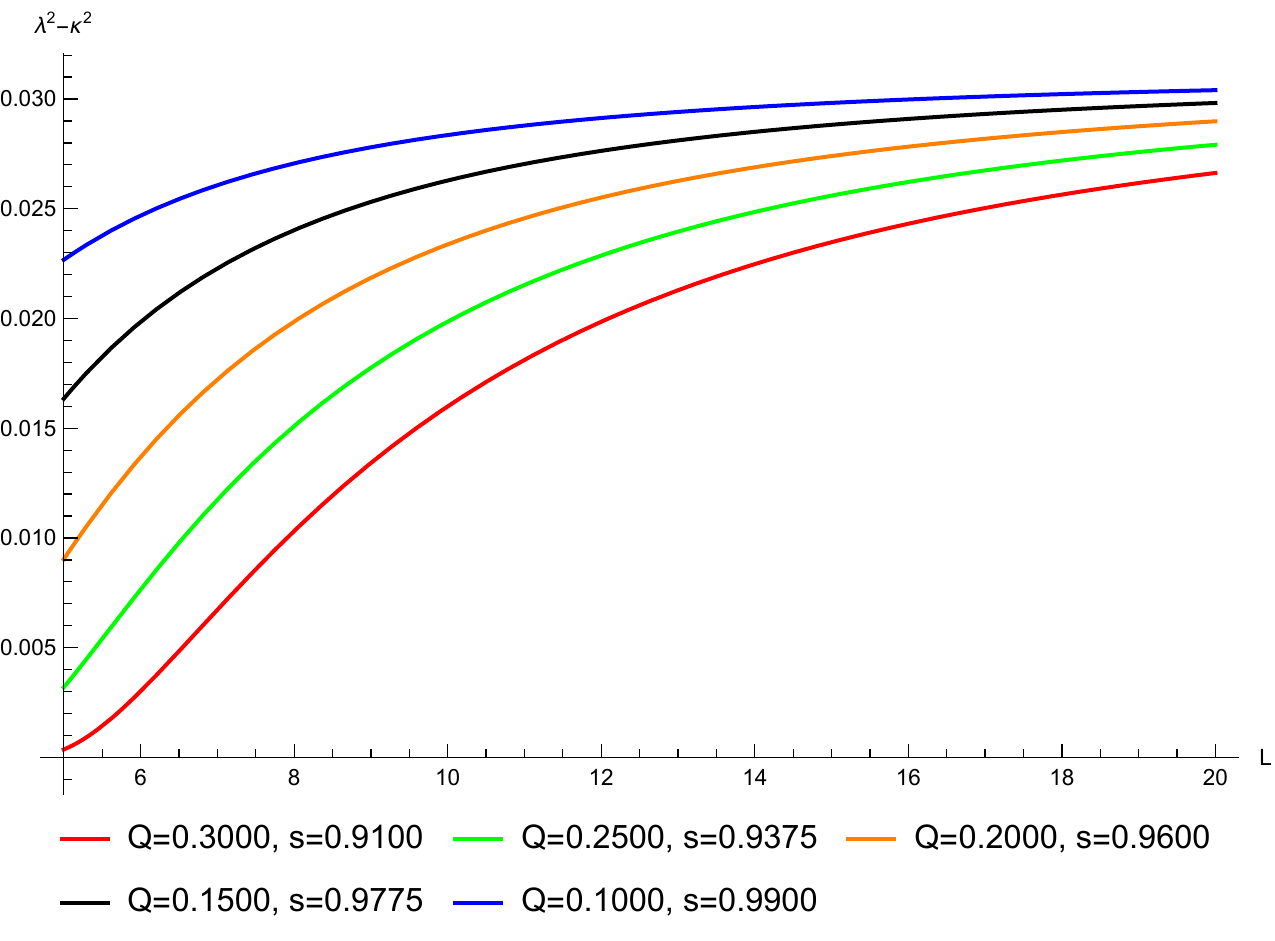}
	\caption{The influence of the angular momentum of the particle around the extremal RNSH black hole on the Lyapunov exponent, where $q=15$. The chaos bound is violated in the range $L>4.39$ ($1.00029r_+<r_0<1.99910r_+$) when $Q=0.3000$ and $ s=0.9100$, in the range $L>3.62$ ($1.00080r_+<r_0<1.99925r_+$) when $Q=0.2500$ and $s=0.9375$, in the range $L>2.83$ ($1.00028r_+<r_0<1.99940r_+$) when $Q=0.2000$ and $s=0.9600$, in the range $L>2.02$ ($1.00110r_+<r_0<1.99955r_+$) when $Q=0.1500$ and $s=0.9775$, and in the range $L>1.12$ ($1.00088r_+<r_0<1.99970r_+$) when $Q=0.1000$ and $s=0.9900$. }
\end{figure}

When the inner and event horizons coincide with each other, the black hole is extremal and the surface gravity disappears. Then there is $Q^2+ s= 1$ and the horizon is located at $r_+ = 1$. To investigate the influences of the charge and angular momentum of the particle around the extremal RNSH black hole on the exponent, we plot Figure 8 - Figure 10.

The influence of the charge on the exponent is plotted in Figure 8. We observe that the range of the particle's charge and spatial region increase with the increase of the hair parameter's value. When the value of the particle's charge decreases, the values of the exponent for different black hole's charges gradually approaches.
Here we find that when the particle's charge takes the certain values, the equilibrium orbits are very close to the horizon. They are closer to the horizon of the extremal black hole than that of the non-extremal black hole. For example, when $Q=0.3000$, $s=0.9100$ and $q\to 16.99$, we get $r_0 \to 1.000206r_+$. The similar case also occurs in Figure 10. In Figure 10, when the value of the hair parameter increases, the range of angular momentum increases, while the spatial range decreases. In Figure 9, when the value of the hair parameter decreases, the range of angular momentum and spatial region increase. The value of the exponent gradually approaches as the angular momentum increases. When $q = 0.9 $, the particle's angular momentum corresponding to the critical value is not much different for different black hole's charges. When the particle's charge is very large, such as $q = 15$, the angular momentum changes significantly. The reason for this phenomenon may be that the particles are subject to different electromagnetic forces from the black hole and the black hole is extremal.

\section{Conclusions and discussions}

In this paper, we investigated the influences of the charge and angular momentum of the particle around the non-extremal and extremal RNSH black holes on the Lyapunov exponent, and found the spatial regions where the chaos bound is violated. For the certain values of the black holes' parameters, the bound is violated in the regions that from the near-horizon regions to the certain distances from the horizons of the non-extremal and extremal black holes. 

The case of the non-extremal black hole was first investigated. For the fixed angular momentum $L=5$, the spatial region for the violation increases with the increase of the black hole's charge when $s=0.20$, and with the increase of the hair parameter's value when $Q=0.70$. For the fixed particle's charges, the angular momentum's ranges and spatial regions increase with the increase of the black hole's charge when $s=0.20$, and with the increase of the hair parameter's value when $Q=0.70$. 
For the extremal black hole, the range of the particle's charge and spatial region decrease when the angular momentum is fixed at $L=5$ and the hair parameter's value decreases. When the particle's charge is fixed at $q=0.9$ and the hair parameter's value decreases, the angular momentum's range increases and spatial region decreases. For the particle with charge $q=15$, there is no obvious trend in the spatial region with the decrease of the hair parameter's value. This may be caused by the special properties of the extremal black hole.

Recently, the weak gravity conjecture has attracted much attention. It asserts that, for the lightest charged particle along the direction of a basis vector in charge space, the charge-to-mass ratio is larger than those for extremal black holes. The ratio of the extremal RNSH black hole is one when the hair parameter is zero. In fact, the value of the hair parameter can be negative \cite{CB}. Therefore, it is feasible that the charge-to-mass ratio of the particle discussed in this paper is greater than 1. In the calculation, we found that the bound is violated in the near-horizon regions of both extremal and non-extremal black holes when the particle's charge is relatively large. For example, the bound is violated at $r_0 \to 1.00463r_+$ when $q=15$, $Q=0.89$, $s=0.20$ and $L \to 4.25$, and at $r_0 \to 1.000206r_+$ when $q \to 16.99$, $Q=0.3000$, $s=0.9100$ and $L =5$. This result is consistent with that obtained in \cite{ZLL,LG2}. In their work, they found that the bound is violated by considering the contributions of the sub-leading terms in the near-horizon expansion. In \cite{LG2}, the authors perceived that this violation should be related to the dynamical stability of the black hole. The study on the stability of the extremal black holes may help us to understand the violation. In \cite{LTW}, the authors taken into account the minimal length effects on the chaotic motion of particles, and found the violation of the bound. They believed that their result does not necessarily implies the violation for the bound conjectured in \cite{MSS}, and the bound should be corrected. In our work, because the backreaction of the particle on the background spacetime was neglected, we got the case that the bound is violated. If the backreaction is taken into account \cite{LG2}, the result may change, which requires the further study.

\acknowledgments
This work is supported by the NSFC (Grant No. 12105031) and Tianfu talent plan.

\end{document}